# Transmission line condition prediction based on semi-supervised learning

Sizhe Li, Xun Ma, Nan Liu, Yi Jin

Nanjing University Of Finance & Economics, Nanjing, Jiangsu, China, 210000

**Abstract**: Transmission line state assessment and prediction are of great significance for the rational formulation of operation and maintenance strategy and improvement of operation and maintenance level. Aiming at the problem that existing models cannot take into account the robustness and data demand, this paper proposes a state prediction method based on semi-supervised learning. Firstly, for the expanded feature vector, the regular matrix is used to fill in the missing data, and the sparse coding problem is solved by representation learning. Then, with the help of a small number of labelled samples to initially determine the category centers of line segments in different defective states. Finally, the estimated parameters of the model are corrected using unlabeled samples. Example analysis shows that this method can improve the recognition accuracy and use data more efficiently than the existing models.

As the main component of the power system, the operation status of transmission lines has an important impact on system safety and stability [1-3]. Some lines are susceptible to defects such as insulation degradation due to long-term influence of environmental factors. Although the general defects do not affect the line operation at the beginning, with the accumulation of time, its severity may continue to increase and cause failure. Therefore, it is necessary to carry out regular inspections and condition assessment of the lines to solve potential safety hazards in time to avoid accidents. Unlike single equipment, overhead transmission lines have a wide span and dispersed condition parameters, so their defects are mainly focused on the overall condition, and the evaluation results often determine the choice of maintenance strategy.

At present, the research related to transmission line condition assessment mainly focuses on the selection of evaluation indexes and the establishment of evaluation methods, while the research on the analysis of line defect-related factors and their prediction methods is relatively small. For example, literature [4] obtained the key parameters by mining the association rules between the line historical defects, fault conditions and the basic parameters, and using the principal component analysis method; literature [5] constructed the transmission line state quantity system and established a classification model based on the random forest algorithm, and optimized the relevant parameters; literature [6] used the fuzzy comprehensive evaluation method to assess the transmission line state, and the weights of the evaluation indexes were reasonably adjusted through the improvement of the affiliation function. The weights of the indicators are reasonably adjusted. In addition, most of the current research ignores the differences between different sections of the line, and the differences between the external environment and the line itself often make the defective state change [7]. In order to improve the operation and maintenance level of transmission lines, it is necessary to synthesize historical data from multiple sources, analyze the relationship between the defect status of each section and different characteristic quantities, and then achieve the status prediction [8]. Differentiated evaluation and accurate state prediction based on relevant characteristic quantities can enable operation and maintenance personnel to focus on the more serious defects, reasonably formulate operation and maintenance strategies, and improve the efficiency of line maintenance [9-10].

Existing prediction models are mainly divided into two categories: traditional algorithms and deep learning algorithms. The former relies more on artificial experience, which is highly subjective, and there are large differences in prediction accuracy. For example, literature [11] selects input features artificially based on the confidence and support of association rules; literature [12] adopts the synthetic minority over-sampling technique SMOTE (Synthetic Minority Over-Sampling) and the decision tree algorithm to predict the state of electric power equipment, in which the amount of the state of each component needs to be selected by humans; literature [13] uses its own components and meteorological factors to predict line failure rates, where the determination of the affiliation functions and weights of the different metrics requires significant expertise. The latter uses models such as deep neural networks in literature [14-15] to autonomously learn the prediction rules through multiple sources of data such as maintenance records, experimental information, and operational parameters, etc. Although the burden of algorithm design is alleviated, the model's demand for data is significantly increased, making it difficult to be replicated.

In this paper, we refer to the evaluation guidelines, use the transmission line multi-source data to construct the historical defect state library; on this basis, we fully consider various defect-related factors, construct the extended feature vector, and use the regular matrix and the representation learning EL (embedding learning) to solve some of the problems of missing data and sparse coding, etc.; for the problems such as poor robustness of the traditional methods and high data demand of the deep learning algorithm, we introduce the semi-supervised learning SSL (semi-supervised learning) technology, use the labelled data to initially obtain the category centre, and then use the unlabelled samples to correct the model estimation parameters. demand, the introduction of semi-supervised learning SSL (semi-supervised learning) technology, the use of labelled data to initially obtain the category centre, and then use the unlabelled samples to correct the model estimation parameters, effectively alleviating the phenomenon of model overfitting; as an example, multiple overhead transmission lines in a certain area are used to validate the superiority of the algorithm proposed in this paper in terms of the accuracy rate, and the efficiency of sample use. The samples are used more efficiently.

| unit | evaluating indicator |
|---|---|
| Basic work and protective facilities | The buried depth of the cable foundation is lower than the designed value, the length of the corrosion diameter of the cable rod is reduced, the tower foundation is pulled |
| pole | Slope, deflection, skew, curvature of the main material, concrete rod crack |
| Guide to the ground | Corrosion, broken strands, damage and flashover and other injuries, sag to the ground distance, ice dancing |
| electrical insulator | Tilt, composite insulator detection, porcelain plate suspension insulator detection, appearance inspection, cleaning, and filth |
| armour clamp | Configuration, protection, connection, crack, discharge gap deviation of ground insulator |
| Lightning protection and grounding device | Breaking base of grounding lead line, unqualified base of grounding resistance, damage diameter of grounding lead line |
| Attached facilities | Defect of rod plate, online monitoring of device defect, damage of bird prevention facilities, climbing ladder guardrail defect, defect of auxiliary communication facilities |
| Channel environment | The crossing distance of the line to all kinds of poles, trees and buildings below, and the buildings and trees in the passage |

**Fig. 1 Assessment system for transmission lines**

# 1 Comprehensive evaluation of the historical state of transmission lines

The comprehensive evaluation of transmission line historical state refers to the relevant evaluation guidelines, the data sources include historical fault and defect records, inspection records, etc., and the evaluation method is to score and synthesise the different unit states. The comprehensive evaluation system shown in Figure 1 is established by the operational characteristics of transmission lines and the studies in the literature [14,16], and the overall index contains two levels. Among them, level 1 is the eight basic equipment units, i.e., foundation and protection facilities, towers, fittings, insulators, conductors, lightning protection and grounding devices, ancillary facilities, and access environment; level 2 is the specific indicators for each basic unit. The scoring steps are as follows.

**Step 1**: With respect to the specific state quantities in FIG. 1, based on the historical records, judge the state degree of each state quantity and deduct points for the corresponding defects.

**Step 2**: Calculate the score of different equipment units from the deduction value of each state quantity and the corresponding weighting coefficient, that is,

$$M_i = \sum_{k=1}^{p_i} w_k^i u_k^i \qquad (1)$$

where: $M_i$ is the score of the $i$ equipment unit after considering all $p_i$ indicators, $i=1, 2, p_i$; $w_k^i$、$u_k^i$ are the weight coefficients and basic demerit points of the $k$ indicator in the $i$ equipment unit, respectively.

**Step 3**: According to the score of each equipment unit, calculate the overall score of the line and judge the overall state of the line section, that is,

$$N = 100 - \sum_{i=1}^{8} \alpha_i M_i \qquad (2)$$

where $N$ is the overall score of the line section after considering 8 equipment units; $\alpha_i$ is the weight coefficient of the $i$ device unit. The above weight coefficients are obtained by the analytic hierarchy process to obtain the judgment matrix, that is, the pairwise comparison method analyzes the importance of different equipment. Through eigenvalue decomposition, normalization and consistency test, the weight vector is finally $[0.062, 0.198, 0.198, 0.110, 0.110, 0.062, 0.062, 0.198]$. Different score intervals correspond to different state levels, of which $(0, 75]$ are serious states; $(75, 85]$ is the abnormal state; $(85, 95]$ is the attention state; $(95, 100]$ is normal.

# 2 Transmission Line Condition Prediction Feature Set Construction

## 2.1 Expanded feature vectors based on multi-source data

Transmission line defect status analysis needs to consider both internal factors and external environment.

The internal factors are the characteristics of the line itself, including voltage level, conductor condition, tower condition, and years of operation. The defective condition of the line varies significantly under different characteristics, so the above information should be included in the condition assessment model. External environment refers to the external factors that can cause defects to occur, mainly meteorological features and temporal and spatial characteristics, including temperature, humidity, wind speed, rainfall, seasons, terrain and zones. The introduction of the above external features can effectively improve the refinement of the evaluation model and achieve the differentiated operation and maintenance of transmission lines. At present, studies on line condition evaluation often only consider the line's own characteristics, while ignoring the influence of environmental factors on the line, so this paper constructs an extended feature vector that introduces external environmental features, that is,

$$x = [s_1, \cdots, s_8, e_1, \cdots, e_6, t_1, \cdots, t_4] \quad (3)$$

where: $x$ is the expanded feature vector; $s_1, \cdots, s_8$ are the characteristics of the line itself (voltage level, number of conductor splits, conductor type, tower call height, full height, stall distance, tower type and years of operation, etc.); $e_1, \cdots, e_6$ are meteorological characteristics, including temperature, humidity, wind speed, rainfall, lightning level and haze level; and $t_1, \cdots, t_4$ are spatial and temporal volume characteristics, including quarterly information, topographic information, elevation, and special section information.

The main sources of the above features include line book information, defect records, spatial information system GIS (geographic information system) and local meteorological data. For the raw data, the quantification rules are as follows: ① Since the own features and the spatio-temporal volume features are relatively fixed over a period of time, different features can be coded hierarchically; ② The meteorological features change more frequently, and since the defects often last for a period of time, it is necessary to comprehensively consider the meteorological features within the period of time, so the raw meteorological data are a series of high-dimensional time sequences. Table 1 gives the meteorological characteristics of a defect record within a certain time period, where the resolution of time is days and $t$ is the current moment. The use of high-dimensional feature vectors under limited sample conditions cannot guarantee the accuracy of the model, and the mode in which the meteorological features are located is more important in transmission line condition assessment than the original data. Therefore, this paper firstly uses principal component analysis to reduce the dimensionality of the original feature vectors, and then uses the K-means method to classify the data, and finally takes the classified pattern information as the model input.

**Tab. 1 Original meteorological data in defect records**

| Characteristics | Time/d | | | | |
| --- | --- | --- | --- | --- | --- |
| | t-4 | t-3 | t-2 | t-1 | t |
| Temperature/°C | 28.4 | 27.6 | 25.4 | 26.4 | 22.3 |
| Relative Humidity/% | 86.2 | 60.3 | 65.3 | 82.3 | 72.6 |
| Wind Speed Rating | 3 | 3 | 1 | 2 | 4 |
| Rainfall/mm | 89.31 | 23.4 | 34.5 | 82.4 | 29.1 |
| Lightning Level | 2 | 0 | 0 | 1 | 0 |
| Haze Rating | 3 | 1 | 2 | 1 | 2 |

**2.2 Missing data filling based on regular matrices**

The expanded transmission line feature vectors $x \in \mathbf{R}^n$, $\mathbf{R}$ are real numbers and $n$ are feature vector dimensions as described in Eq. (3), which need to be further processed. In practical application scenarios, some missing features are difficult to avoid, so this paper adopts a regular matrix-based complementation strategy [17]. For the original matrix composed of feature vectors, the low rank decom-position is used to obtain the approximate matrix as the filling value, and the core idea of this method is to obtain the best approximate matrix as the input of the subsequent model by optimally iterating the objective function, that is,

$$X \approx UV = \hat{X} \quad (4)$$

$$J = \sum_{i=1}^{m}\sum_{j=1}^{n} \| X_{ij} - U_i^\mathrm{T} V_j \| + \frac{\lambda}{2}\left(\| U \|_2^2 + \| V \|_2^2\right) \quad (5)$$

where: $X = [x_1^\mathrm{T}, \cdots, x_i^\mathrm{T}, \cdots, x_m^\mathrm{T}]^\mathrm{T}$ is the original feature matrix; $x_i$ are the expanded eigenvectors $i=1,2,\cdots,m$ corresponding to the $i$ cases; $m$ is the total number of cases; $U$、$V$ are the low-rank matrices obtained from decomposition; $U_i$ is the $i$ columns of the $U$ matrix; $V_j$ is the $V$ columns of the $j$ matrices; $\hat{X}$ is the approximation matrix; $J$ is the optimization objective function; $X_{ij}$ is the known eigenvalues of the original feature matrix; $\lambda$ is the regular term coefficients. For the missing quantities in the original feature matrix, the elements in the corresponding positions of $\hat{X}$ are used as substitutes, and the expanded eigenvectors are written as $\hat{x}$ after filling.

**2.3 EL-based feature mapping**

Expanded feature vectors are essentially high-dimensional discrete random vectors, where each dimension represents the category in which different feature quantities are located. Under the condition of limited samples, the direct use of the sparse coding method described in Section 2.2 cannot guarantee the computational efficiency and prediction accuracy, so EL is introduced in this paper [18]. The core idea of EL is to transform the original high-dimensional discrete vectors into low-dimensional continuous features by using the multilayer perceptron (MLP), that is,

$$v^{l+1} = f_\sigma(W^l v^l + b^l) \quad (6)$$

Where: $v^l$ is the feature vector of layer $l$; $f_\sigma$ is the activation function; $b^l$ and $W^l$ are the bias vector and weight matrix of layer $l$, respectively. For the transmission line state prediction problem, the input to the bottom layer of the model is the expanded feature vector after filling, that is, $v^0 = \hat{x} \in \mathbf{R}^{18}$, and the final output features are $v^L \in \mathbf{R}^6$ after the MLP model with $L$ layers.

## 3 Transmission line condition prediction under semi-supervised conditions

The core idea of transmission line state prediction is to mine the development pattern of line defects based on historical data, and then predict the line state in the future moment. The essence of the clustering problem, that is, different modes of historical data is divided into different categories, while the same mode of the line in the future moment is often the same state. The basis of clustering is similarity calculation, which is more reasonable and efficient than the method of calculating the similarity between different samples and the current sample, which first obtains the category centre and then calculates the similarity between different category centres and the current sample. In practical application scenarios, data annotation is time-consuming and labour-intensive, and often requires the guidance and help of professionals. Therefore, one of the key issues in this paper is how to reasonably use unlabelled data to improve the model prediction accuracy, i.e., SSL[19]. The core idea of the algorithm proposed in this paper is to first determine the initial clustering centre using the labelled data, and then judge the corresponding category of the unlabelled data according to the clustering centre, that is, let the number of all training samples be $N$, where the number of labelled samples is $N_s$, and the number of unlabelled samples is $N_q$. The sample point set consists of labelled points $(v_{s,m}, y_m)$, $m \in \{1,\cdots,N_s\}$; unlabelled points $v_{q,n}$, $n \in \{N_s+1,\cdots,N\}$, and $y_m \in \{1,2,3,4\}$ are the overall status of the zone, which are normal, attention, abnormal and serious respectively.

According to the labeled data, the class center corresponding to the overall state of different sections can be determined, that is,

$$c_k = \frac{1}{|S_k|} \sum_{m' \in \{m: y_m = k\}} v_{s,m'} \quad (7)$$

where: $c_k$ is the clustering centre corresponding to the overall state category $k$ of the zone; $S_k$ is the set of category $|S_k|$ of the labelled samples; $k$ is the set size. The four category centres are $c_1$、$c_2$、$c_3$、$c_4$ and $(v_{s,m}, y_m)$ for the set of labelled points.

Unmarked points are classified using the category centre,

$$\hat{z}_n = \arg\max_k P(z_n = k | v_{q,n}) \quad (8)$$

$$P(z_n = k | v_{q,n}) = \frac{\exp(-d(v_{q,n}, c_k))}{\sum_{k'=1}^{4} \exp(-d(v_{q,n}, c_{k'}))} \quad (9)$$

$$d(v_{q,n}, c_k) = \| v_{q,n} - c_k \|_2 \quad (10)$$

where: $\hat{z}_n$ is the prediction obtained based on the labelled samples; $P$ is the conditional probability; $z_n$ is the overall state of the zone corresponding to the unlabelled samples $v_{q,n}$; $c_k$ is the clustering centre corresponding to the categories $k$; $d$ is the Euclidean distance.

So far, from the prediction results of the overall state of the zones of the unlabelled data, the clustering centre can be adjusted so as to overcome the problem of the low reliability of the estimation results of the category centre under the condition of small samples. Considering that the predicted labels of the unlabelled data are used and the confidence parameter $\alpha$ is introduced, the corrected category centre $c'_k$ can be expressed as

$$c'_k = \alpha c_k + (1-\alpha) \frac{1}{|S'_k|} \sum_{n' \in \{n: \hat{z}_n = k\}} v_{q,n'} \quad (11)$$

Where: $c_k$ is the centre of the category corresponding to the labelled samples; $S'_k$ is the set of predicted categories $k$ for the unlabelled samples; $|S'_k|$ is the set size. The confidence level $\alpha$ can be determined by cross-validation, and in this paper it takes the value of 0.15.

At this point the SSL is complete, and for the new sample $v_x$, the category can be judged in the following manner

$$P(h_x = k | \boldsymbol{v}_x) = \frac{\exp(-d(\boldsymbol{v}_x, \boldsymbol{c}'_k))}{\sum_{k'=1}^{4} \exp(-d(\boldsymbol{v}_x, \boldsymbol{c}'_{k'}))} \tag{12}$$

where: $P(h_x = k | \boldsymbol{v}_x)$ is the probability that the overall state of the zone belongs to category $k$; $h_x$ is the overall state of the zone corresponding to sample $\boldsymbol{v}_x$.

The cross-entropy function is used for model training, and the visualisation results of SSL category centre estimation are given in Fig. 2, where the visualisation is based on principal component analysis, with the horizontal coordinate PC1 denoting the first principal component and the vertical coordinate PC2 denoting the second principal component in the figure, and PC1 and PC2 are normalised. In Fig. 2(f), the hollow and solid points indicate the results under supervised and semi-supervised conditions, respectively, where the results under supervised conditions are the category centres corresponding to the labelled samples; the results under semi-supervised conditions are the adjusted results after adding the unlabelled samples. From Fig. 2(f), it can be seen that the category centre is corrected after adding unlabelled samples, which is conducive to improving the accuracy of the model under small sample conditions. Figure 3 gives an overall schematic of the methodological framework proposed in this paper.

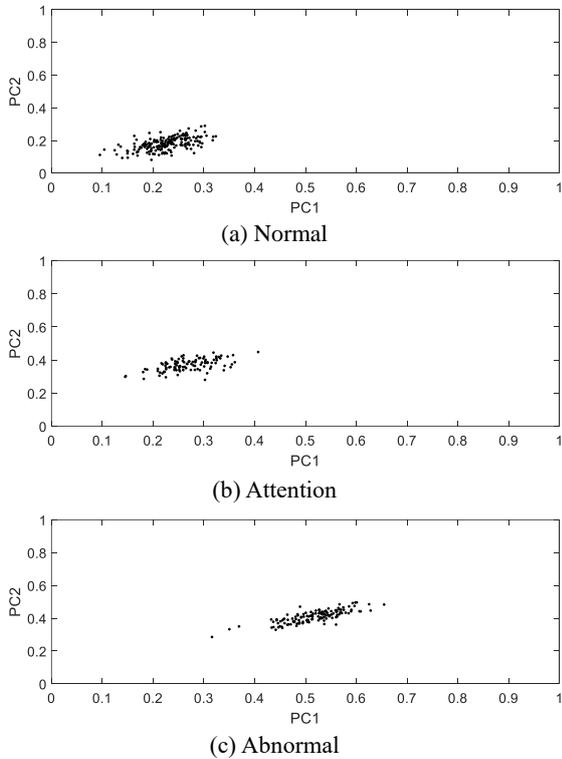

(a) Normal

(b) Attention

(c) Abnormal

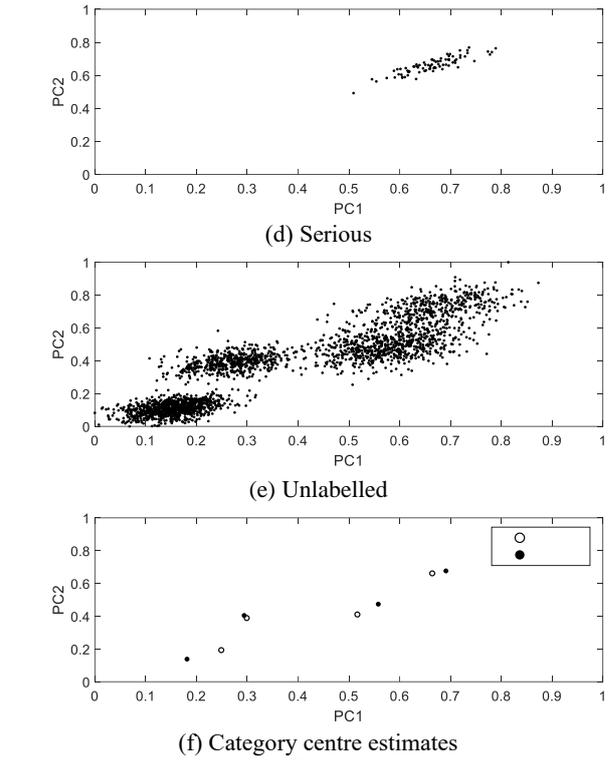

(d) Serious

(e) Unlabelled

(f) Category centre estimates

**Fig. 2** Visualization result of class center estimation in semi-supervised learning

## 4 Case analysis

### 4.1 Expanded feature vectors based on multi-source data

The algorithm uses multi-source data of overhead transmission lines in a region from July-October 2019, which contains 2,250 records, each of which corresponds to the full amount of features in a single week for a single line segment, of which the proportion of records with partial feature deficiency problems is 29.3%. Firstly, the steps and methods in Section 1 are used to evaluate the overall state of the line in each section, i.e., after the extraction and preprocessing of the defective conditions, the defective conditions are scored and synthesised according to the evaluation indexes under different equipment units; then, the number of the labelled data is 1,000, of which some of the evaluation results are shown in Table 2, and the distribution of the samples of different defective conditions is given in Table 3, in which half of the labelled data is used as the training samples, and the rest of the labelled data are used for testing; finally, a total of 1,250 pieces of unlabelled data are used to test the effect of the SSL strategy proposed in this paper on the improvement of prediction accuracy.

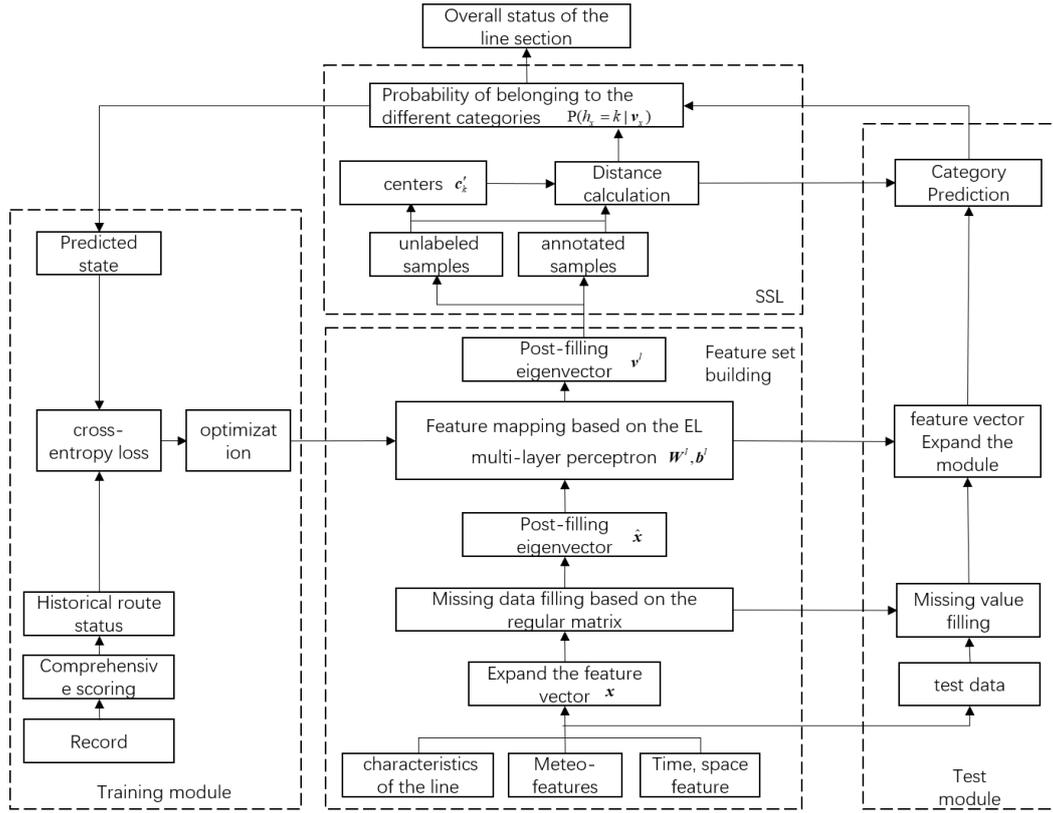

**Fig. 3 Schematic of the proposed framework**

**Tab. 2 Overall evaluation results of some overhead line sections**

| Zones | Deficiencies | Rating | Grade |
|---|---|---|---|
| T3 | Anti-vibration hammer corrosion Corrosion of fittings, mechanical strength reduced to 87%. Damage to the right upper subconductor on the large-sized side of the right phase, with a depth of damage of 15 per cent | 88.4 | Attention |
| T7 | Some of the tower bolts are corroded Slight corrosion of tie rods Small loss of foundation retaining soil | 92.1 | Attention |
| T12 | C-leg main material bending degree 0.25 per cent 3% bending of connecting device crimp tube Minor corrosion of tower bolts Loose or missing bolts in 8% of tower components | 80.7 | Abnormal |

**Tab. 3 Distribution of samples in different defect states**

| Sample type | Sample size | | | | total |
|---|---|---|---|---|---|
| | Normal | Attention | Abnormal | Serious | |
| Ttraining sample | 288 | 126 | 58 | 17 | 500 |
| Test sample | 320 | 114 | 49 | 28 | 500 |
| Unmarked sample | Unknown | Unknown | Unknown | Unknown | 1250 |

The current evaluation metric commonly used for classification tasks is the F1 score, which is calculated as follows:

$$F1 = \frac{2}{\alpha_r^{-1} + \alpha_p^{-1}} = \frac{N_{tp}}{N_{tp} + \frac{1}{2}(N_{fp} + N_{fn})} \quad (13)$$

Where: $\alpha_r$ and $\alpha_p$ are recall rate and accuracy rate respectively ; $N_{fp}$, $N_{fn}$ and $N_{tp}$ were the number of false positive samples, false negative samples and true positive samples, respectively.

Compared to accuracy, the metric also considers recall. For conservative models, the accuracy can be improved by screening simple samples and ignoring difficult samples. While the F1 score serves as a reconciled average of the accuracy and recall, the F1 score increases significantly only when both are at a high level. In addition, the problem of category imbalance is well avoided because the calculation method is to calculate the F1 score separately for each category and then average it.

### 4.2 Analysis of results based on the proposed SSL framework

Firstly, we quantified the feature quantities of the complete data, among which, the self-features and spatio-temporal features were coded by hierarchical coding; and the meteorological features were coded by pattern coding.

The pattern coding consists of principal component analysis and K-means clustering, and the number of principal components is set to 4 and the number of clustering categories is set to 5. For some of the missing feature quantities, the filling strategy based on the regular matrix is adopted, and the coefficient of the regular term in the optimization objective is set to 0.1. For the expanded feature vectors after the filling process, the MLP is used to perform the EL, where the number of MLP implicit layers is 1, the number of neurons is 12, the activation function is the ReLU function, the dimension of the input layer is 18, the dimension of the output layer is 18, and the dimension of the output layer is 12. 12; the activation function is ReLU function; the dimensions of the input layer are 18 and the dimensions of the output layer are 6. The overall state prediction of the zones can be achieved based on the representation vectors, where the centre of each category is the average value of the representation vectors under that category. For labeled data, the corresponding category of the representation vector is known; for unlabeled data, the category is predicted based on the category centre, and then the centre position is corrected using the representation vector until the model converges.

The prediction results under supervised learning and semi-supervised learning conditions are given in Tables 4 and 5, respectively. As can be seen from Tables 4 and 5, the overall F1 score increases by 16.2% under semi-supervised conditions, and the error results have less deviation from the actual state, in the case of limited number of samples, the correction of unlabelled samples to the centre of the category is extremely critical, which can effectively alleviate the overfitting phenomenon; for the F1 scores of each category, the line segment with abnormal defective state has the largest improvement under SSL conditions, which is because the samples under this state are fewer and easily confused with other states. This is due to the fact that the samples in this state are small and easily confused with other states, and the unlabelled data can further differentiate the cha-racterisation vectors.

**Tab. 4 Prediction results under supervised learning**

| Actual status | Projected state | | | | F1 Score | Overall F1 score |
|---|---|---|---|---|---|---|
| | Normal | Attention | Abnormal | Serious | | |
| Normal | 287 | 25 | 8 | 0 | 0.924 | |
| Attention | 14 | 85 | 15 | 0 | 0.720 | 0.723 |
| Abnormal | 0 | 12 | 29 | 8 | 0.532 | |
| Serious | 0 | 0 | 8 | 20 | 0.714 | |

**Tab. 5 Prediction results under semi-supervised learning**

| Actual status | Projected state | | | | F1 Score | Overall F1 score |
|---|---|---|---|---|---|---|
| | Normal | Attention | Abnormal | Serious | | |
| Normal | 298 | 18 | 4 | 0 | 0.949 | |
| Attention | 10 | 102 | 2 | 0 | 0.857 | 0.885 |
| Abnormal | 0 | 4 | 43 | 2 | 0.843 | |
| Serious | 0 | 0 | 4 | 24 | 0.889 | |

In addition, correlation analyses allow the identification of critical features in different defect states. In the line section under the attention state, special section information such as temperature, rainfall, wind speed class, quarterly information, and terrain information are more critical; in the line section under the abnormal state, special section information such as voltage class, tower type, commissioning age, rainfall, lightning class, haze class, and quarterly information are more critical; and in the severe state, the amount of the line characteristics exists a strong randomness, and no obvious causative factors are found.

### 4.3 Compare and contrast common classification models

Some of the commonly used classification models compared in this paper are Bayesian network BN (Bayesian network), support vector machine SVM (support vector machine) model and long short-term memory LSTM (long short-term memory) network. Among them, BN is based on the probability calculation of directed acyclic graph, and solves the uncertainty problem of transmission line state evaluation by constructing the causality graph between evaluation indexes; LSTM network simulates the human memory mechanism to mine the input vectors through the forgetting gate and the reset gate; SVM uses the kernel function to map the input vectors to the linearly divisible intervals, and divides the feature points by using the hyperplane. The specific implementation details are as follows: the a priori probability in BN is set in reference [16], the conditional probability table CPT (Conditional Probability Table) learning is obtained by using the training dataset of this paper, and the learning method of the network structure adopts the K2 scoring algorithm; the inputs of the LSTM network model are the affiliation values of the different metrics, and the affiliation function is set in reference [14], the number of LSTM network layers is 3, the number of units in each layer is 128, 64, 32, respectively, the 3rd layer is set to have a random deactivation rate [20] of 0.2 and a learning rate of 0.0001; the input of the SVM model is the expanded feature vector, a polynomial kernel function is used, and its output is the overall state of the line segment.

**Tab. 6 Comparison of result among different methods**

| Defect Status | BN | LSTM | SVM | SSL |
|---|---|---|---|---|
| Normal | 0.745 | 0.811 | 0.803 | 0.949 |
| Attention | 0.640 | 0.739 | 0.742 | 0.857 |
| Abnormal | 0.603 | 0.666 | 0.682 | 0.843 |
| Serious | 0.672 | 0.693 | 0.705 | 0.889 |
| Overall | 0.665 | 0.727 | 0.733 | 0.885 |

The comparison results of different methods are shown in Table 6, from which it can be seen that the SSL of the proposed method is much better than the remaining three models. This is because the proposed method makes the parameter estimation more reasonable under small samples with the help of unlabelled samples; in addition, the means of filling the missing data based on the regular matrix, representation learning and category-centred updating strategy considering the confidence level have effectively optimized the model learning process, which further widens the performance gap between different models.

Considering the cost of data collection and labelling, the practical application of engineering is more sensitive to the amount of labelled data, Fig. 4 gives the change of F1 scores in the test set under different numbers of labelled samples. Keeping the test set unchanged during the experiment, the labelled data in the training set is gradually transformed into unlabelled data. As can be seen from Fig. 4, the method proposed in this paper is much less affected by the number of labelled data than the remaining three methods, and the test accuracy is obviously at a higher level. In summary, the method proposed in this paper is more suitable for the task of transmission line state prediction in real scenarios.

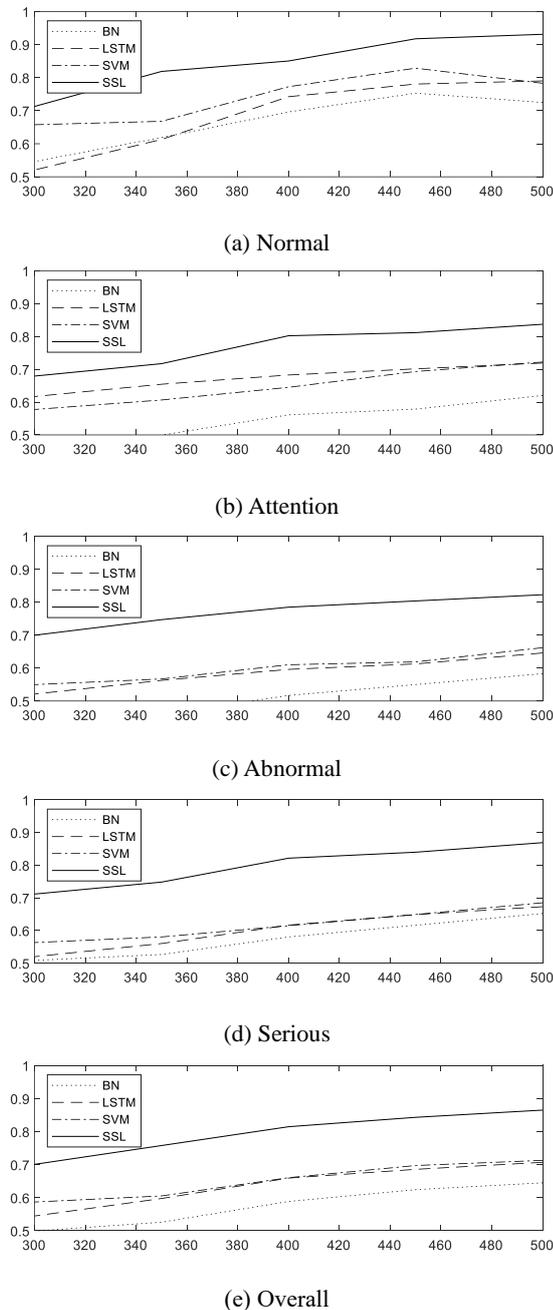

(a) Normal

(b) Attention

(c) Abnormal

(d) Serious

(e) Overall

**Fig. 4 Curves of F1 score vs the number of labelled samples in training set**

# 5 Conclusion

In this paper, we propose an SSL-based overhead transmission line condition prediction method, which integrates multi-source data to provide a basis for operation and maintenance decision-making. The main conclusions are as follows.

(1) Considering multi-source information comprehensively, obtaining the historical defective status of transmission lines based on evaluation guidelines, and using improved hierarchical analysis to get the weights of different equipment units.

(2) Construct the extended feature vector containing its own features, meteorological features and spatio-temporal features, and adopt the strategy based on the regular matrix to fill in the missing data, and improve the operation efficiency and prediction accuracy with the help of EL.

(3) Use the labelled samples in the representation space to obtain the category centre when the line segment is in different states, and introduce the unlabelled samples to correct the parameter estimation results on the basis of this, so as to achieve SSL under a small number of labelled samples. Compared with the existing methods, the accuracy of the model proposed in this paper is improved by about 16.2%, which provides a new way of thinking for the state assessment of key equipment in the electric power industry.

# Reference：